# High-Entropy Oxide Nanostructures for Rapid and Sustainable Nitrophenol Reduction

Anjali Varshney[a], Aishwery J. Verma[b], Ritesh Dubey[a], Sushil Kumar[a,*], Tapas Goswami[a,*], and Samar Layek[b,*]

High-entropy materials have emerged as promising class of catalysts, driven by their high configural entropy originated from structural disorders in single phase multi-component systems. Despite their potential, catalytic performance of high-entropy oxides (HEO) are relatively less explored. In this study, we present a simple solution-based combustion route to synthesize two low-cost, transition metal-rich multicationic oxides positioned in the medium- (**HEO-4**) and high-entropy (**HEO-5**) region. Rietveld refinement of the powder X-ray diffraction data confirmed single phase formation and *fcc* crystal structure in both these nanostrucres. Morphology, size and multicationic elemental analysis were investigated using both scanning and transmission electron microscopic studies. The catalytic performance of these HEOs was evaluated in the hydrogenation of a series of nitrophenol derivatives. Notably, **HEO-5** displayed significantly higher catalytic activity ($k_{app} \approx 0.5$ min$^{-1}$, TOF = $2.1 \times 10^{-3}$ molg$^{-1}$s$^{-1}$), achieving rapid conversion of *p*-nitrophenols compared to medium-entropy oxide nanostructures ($k_{app} \approx 0.02$ min$^{-1}$, TOF = $7.2 \times 10^{-4}$ molg$^{-1}$s$^{-1}$). Furthermore, the reaction kinetic and thermodynamic parameters ($E_a$, $\Delta H^\ddagger$, $\Delta G^\ddagger$ and $\Delta S^\ddagger$) were determined in order to gain mechanistic insight into the reduction process. This study opens avenues for developing rational design and facile synthetic strategies for HEOs as efficient catalysts toward large-scale sustainable amine production.

## Introduction

High-entropy (HE) materials have recently emerged as a versatile platform for sustainable catalytic transformations.[1–4] Initially developed for advanced metallic alloys, the high-entropy concept was subsequently extended to oxide systems, giving rise to high-entropy oxides (HEOs), which are multicationic complex ceramic systems.[5,6] These oxides exhibit distinctive physicochemical properties arising from configural entropy stabilization. By incorporating multiple metal cations (typically five or more in near-equiatomic ratios) into a single-phase crystal lattice, HEOs often display exceptional and sometimes unpredictable functional behavior.[7] The entropy-driven stabilization effect facilitates the formation of simple solid-solution structures, even in the presence of numerous competing cations, thereby enhancing both structural stability and electrochemical performance.[8]

HEOs exhibit superior properties compared to conventional mixed-metal oxides, including exceptional structural stability, improved electrochemical performance, and high thermal and chemical durability.[9] They also demonstrate efficient photothermal conversion[9] and intriguing magnetic characteristics such as superconductivity, soft magnetism, and superparamagnetism.[10] These unique attributes position HEOs as promising candidates for a wide range of applications, including catalysis, energy storage, and photocatalytic hydrogen generation.[2,8,11–13]

Several synthetic strategies have been employed for the preparation of HEAs/HEOs, including conventional casting, mechanical alloying, plasma spheroidization, laser cladding, magnetron sputtering, and electrochemical deposition.[14,15] However, their practical applications remain constrained by the need for straightforward, scalable, and cost-effective synthetic routes. Among the various HEO families, 3d transition metal-based systems (e.g., Cu, Cr, Mn, Fe, Co, Ni, Zn) have been most widely explored due to their rich redox chemistry, entropy-driven stabilization, enhanced electrochemical behavior, and tunable functional properties. Recent studies have revealed the multifunctional catalytic potential of HEOs across diverse transformations, including oxidation, hydrogenation, and electrocatalysis. Rost *et al.* first reported the entropy-stabilized oxide (MgCoNiCuZn)O in 2015, highlighting its structural stability at elevated temperatures.[16] Subsequently, Sarkar *et al.* demonstrated that equiatomic multicomponent oxides (Mg, Co, Ni, Cu, Zn)O exhibit improved oxygen evolution reaction (OER) activity, arising from synergistic cationic effects.[6] Likewise, Bérardan *et al.* explored transport and defect properties in HEOs, demonstrating how configurational disorder influences ionic conductivity, a factor highly relevant to redox-based catalysis.[17]

In heterogeneous catalysis, Chen *et al.* reported multi-cationic high entropy oxide as efficient catalysts for CO oxidation, attributing their superior activity to the presence of multivalent redox centres and high surface area.[18] Similarly, Shi *et al.* investigated HEOs for $CO_2$ photomethanation reaction, highlighting the role of lattice distortions in accelerating reaction kinetics.[19] More recently, Nguyen *et al.* demonstrated that high-entropy oxide photocatalysts exhibit excellent performance in ammonia synthesis, where entropy-driven stabilization of multiple transition metals generated highly

[a.] Department of Chemistry, Applied Sciences Cluster, UPES Dehradun, Energy Acres Building, Dehradun- 248007, Uttarakhand, India.
[b.] Department of Physics, Applied Sciences Cluster, UPES Dehradun, Energy Acres Building, Dehradun- 248007, Uttarakhand, India.
* Corresponding authors: sushilvashisth@gmail.com, tapas.t@gmail.com, samarlayek@gmail.com

active and stable catalytic centres.[20] Collectively, these studies highlight the unique advantages of HEOs such as structural stability, electronic tunability, and synergistic multicationic interactions, establishing them as a promising class of catalysts for energy and environmental applications.

Despite a significant progress, the application of HEOs in organic transformations remains limited, with the catalytic hydrogenation of nitroaromatic compounds receiving rare attention. The reduction of nitro groups to amines is particularly important, as amines serve as key intermediates in pharmaceuticals, dyes, and agrochemicals, while the reaction itself is widely employed as a benchmark for evaluating catalytic efficiency.[21] Conventional noble-metal catalysts, however, are hindered by high cost, poor stability, and limited recyclability.[22] To overcome these limitations, sustainable catalytic materials have been explored, for instance, electronic waste derived gold nanoparticle[23] have exhibited superior catalytic activity for nitrophenol reduction while transition metal oxide-based catalyst such as CuO NPs have also been employed to drive the same reaction [24]. Recently bimetallic[25] and trimetallic nanostructures[26] have demonstrated enhanced catalytic efficiency towards nitrophenol reduction which is attributed to the synergistic effect and improved electronic interactions. Though few studies have shown that HEAs can effectively catalyse the reduction of nitrophenols,[27–29] comparable investigations on HEOs are scarce. Moreover, the fundamental relationship between their crystal structure, surface chemistry, and catalytic performance remains poorly understood. Addressing this gap forms the core motivation of the present study.

Herein, we present a strategy for the preparation of single-phase HEOs with four-element (Ni, Cu, Co, Zn) and five-element (Ni, Mg, Cu, Co, Zn) compositions. These oxides were synthesized *via* a facile solution combustion synthesis (SCS) method, offering cost-effectiveness, scalability, and molecular-level homogenization of precursors. Phase formation was achieved through high-temperature annealing and confirmed by comprehensive spectro-analytical characterization. The catalytic performance and thermodynamic parameters were further evaluated for the reduction of nitrophenol derivatives to their corresponding amines. The key highlights of this work are 3-fold: (i) a simple and scalable synthesis of HEOs *via* the SCS approach, (ii) the demonstration of HEOs as efficient heterogeneous catalysts for sustainable amine production and as reusable alternatives to noble-metal-based systems, and (iii) the first systematic evaluation of the influence of configurational entropy on catalytic hydrogenation efficiency.

## Experimental Section

### Materials and methods

Glycine, sodium borohydride (NaBH$_4$) and metal nitrates [i.e., Ni(NO$_3$)$_2$·6H$_2$O, Mg(NO$_3$)$_2$·6H$_2$O, Cu(NO$_3$)$_2$·3H$_2$O, Zn(NO$_3$)$_2$·6H$_2$O, and Co(NO$_3$)$_2$·6H$_2$O] were obtained from Merck, while nitroaromatic compounds used in the catalytic reduction studies were procured from SRL Chemicals. All chemicals were used as received without any further purification. Powder X-ray diffraction (PXRD) patterns of HEOs were collected on a Bruker D8 Advance ECO diffractometer equipped with Cu-K$\alpha$ radiation ($\lambda$ = 1.5406 Å) to investigate phase formation and crystal structures of the as-synthesized materials. Transmission electron microscopic (TEM) images were collected using a JEOL electron microscope operating at 200 kV, while field emission scanning electron microscopy (FE-SEM) for morphological analysis was performed on a JEOL JSM-7800F Prime microscope. X-ray photoelectron spectroscopy (XPS) measurements were carried out on a PHI 5000 VersaProbe III instrument using monochromatic Al K$\alpha$ radiation. Functional groups present in the as-synthesized nanostructures were analyzed using a PerkinElmer FTIR spectrometer. Surface area and porosity were determined by Brunauer-Emmett-Teller (BET) analysis using an Anton Paar Autosorb 6100 analyzer. UV-Vis absorption spectra were recorded with a Shimadzu UV-1800 spectrophotometer.

### Syntheses of HEOs

High-entropy oxide nanostructures were synthesized using a solution combustion method. For the five-element composition, (NiMgCuZnCo)$_{0.2}$O (**HEO-5**), equimolar quantities (0.2 M each) of nickel nitrate hexahydrate, magnesium nitrate hexahydrate, cupric nitrate trihydrate, zinc nitrate hexahydrate, and cobalt nitrate hexahydrate were dissolved in 100 mL of deionized water and stirred at 80 °C for 50 min. Glycine ($\approx$0.72 M) was then added as a fuel, and the mixture was further stirred until a homogeneous gel was formed. The gel was subsequently heated to 180 °C, initiating a self-sustained combustion reaction that yielded a loose, ash-like precursor powder. The resulting powder was finely ground, transferred to a quartz crucible, and calcined in a muffle furnace at 100 °C to obtain phase-pure **HEO-5**. The medium-entropy oxide, (NiCuZnCo)$_{0.25}$O (**HEO-4**), was synthesized following the same procedure using four metal nitrates.

### Catalytic activity of HEOs.

The catalytic efficiency of the as-prepared multicationic metal oxide nanostructures (**HEO-5** and **HEO-4**) was evaluated for the reduction of various nitrophenol derivatives, including *p*-NP, *o*-NP, *m*-NP, 2,4-DNP, and PA. In a typical experiment, 30 mL of an aqueous solution of the nitrophenol derivative (10$^{-4}$ M) was mixed with 7.0 mg of NaBH$_4$. The reduction reaction was initiated by adding 0.5 mg of **HEO-5** or **HEO-4** catalyst to the mixture, followed by continuous stirring. The reaction progress was monitored by recording UV–Vis spectra of aliquots collected at regular time intervals.

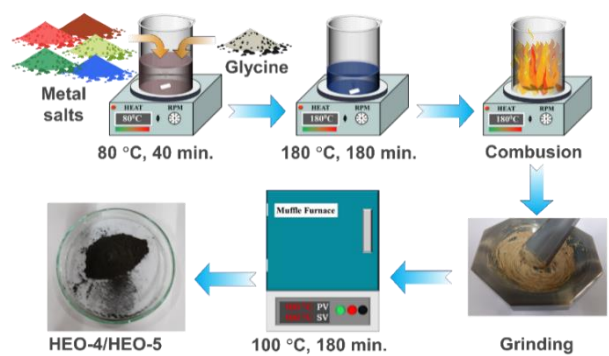

**Scheme 1.** A schematic showing the preparation method used for high-entropy oxides (**HEO-4** and **HEO-5**) in the present study.

## Results and discussion

High entropy oxide (**HEO-5**) was synthesized *via* a facile and scalable solution combustion method[30,31] using metal nitrates as precursors and glycine as the fuel. In this method, the metal nitrates act as oxidizing agents as well the source of the metal ions. Glycine plays a dual role: (i) complexing agent that coordinates with the metal ions and (ii) act as a fuel to drive the combustion reaction.[32,33] The highly exothermic combustion reaction facilitates the uniform distribution of metal ions in the multicationic oxide lattice.[34–37] Additionally, the rapid release of gases during the during the combustion reaction of nitrates leads to the creation of highly porous metal oxide. The overall reaction can be represented as follows:

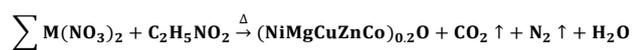

$$\sum M(NO_3)_2 + C_2H_5NO_2 \xrightarrow{\Delta} (NiMgCuZnCo)_{0.2}O + CO_2 \uparrow + N_2 \uparrow + H_2O$$

where, M represents a transition-metal (Ni, Mg, Cu, Zn, Co). The high temperature generated during combustion facilitates solid-solution formation of the multicationic oxide and crystallization in a single phase, stabilized by high configurational entropy. This synthesis route thus offers an energy-efficient, low-cost, and scalable method for preparing high-entropy oxides with homogeneous elemental distribution and porous morphology, which are beneficial for catalytic applications.

### Structural studies

The crystal structures of the high entropy oxides were examined using powder X-ray diffraction study. Both **HEO-4** and **HEO-5** exhibited similar PXRD patterns (Fig. S1), suggesting the formation of a single-phase solid solution. The diffraction peaks observed at 2θ values near 36°, 43°, 62°, 72° and 80° could be indexed to the (111), (200), (220), (311) and (222) crystal planes of *fcc* structure.[16,38] To obtain the structural parameters, Rietveld refinements of the PXRD data was performed using GSAS-II software package.[39] The refinement profiles (Fig. 1(a-b)) revealed excellent agreement between experimental and calculated data when modeled in the *Fm-3m* space group, confirming the formation of a single-phase *fcc* structure corresponding to NiO. The extracted lattice parameters for all investigated phases are consistent with previously reported values for NiO-based solid solutions, further validating the successful synthesis of high-entropy oxides. PXRD Rietveld refinements of **HEO-4** (a = b = c = 4.240 Å) and **HEO-5** (a = b = c = 4.228 Å) indicate a slight variation in lattice parameters.[38] This difference can be correlated with the average ionic radii of the constituent cations.[40] Further analysis of the PXRD data was conducted using the Williamson-Hall (W-H) method to estimate the crystallite size (D) and lattice strain (ϵ). In this method, the total peak broadening ($\beta_{total}$) is expressed as a function of diffraction angle according the relation: $\beta_{total}\cos\theta = \frac{K\lambda}{D} + \varepsilon\, 4\sin\theta$, where $K$ is the shape factor and $\lambda$ is the X-ray wavelength.[41] From the linear plot of $\beta_{total}\cos\theta$ versus $4\sin\theta$ plot (Fig. 2), average crystallite size and micro strain (ϵ) was calculated from the intercept and slope, respectively. The W-H plots revealed that the crystallite size of **HEO-5** (D≈20 nm) was slightly smaller than that of **HEO-4** (D≈25 nm), indicating enhanced lattice distortion with the addition of more metal ions. Moreover, the calculated strain was found to be higher in **HEO-5** compared to **HEO-4**. This increased strain can be ascribed to the larger lattice distortion and local structural disorder introduced by the incorporation of five different cations of varying ionic radii into the single-phase *fcc* lattice.[19,42] This enhanced microstrain in high-entropy metal oxides can potentially lead to enhanced catalytic activity.[13,19]

The morphology of the as-prepared high-entropy oxide nanostructures was examined using electron microscopy, and the results are presented in Fig. 3. SEM micrographs of **HEO-5** and **HEO-4** reveal a highly porous architecture composed of uniformly distributed nanoparticles (Fig. 3a). TEM analysis further confirms the formation of spherical nanoparticles (Fig. 3b). Particle size distribution, determined using *ImageJ* software, indicates an average diameter of ~19 nm for **HEO-5** (Fig. S3). High-resolution TEM (HRTEM) images show clear lattice fringes with an interplanar spacing of 0.243 nm, corresponding to the (111) plane of NiO (Fig. 4c). The selected area electron diffraction (SAED) patterns (Fig. 4d) exhibit diffused rings with bright spots indexed to the (111), (200), (220), and (311) planes, confirming the polycrystalline nature of the nanoparticles, consistent with PXRD and Rietveld refinement results (*vide supra*). Elemental mapping by EDS confirms uniform dispersion of all the constituent metal ions (Ni, Mg, Fe, Cu, and Zn in **HEO-5**, and Ni, Fe, Cu, and Zn in **HEO-4**), with no detectable impurities (Figs. S4 and S5). EDAX spectra from both SEM and TEM analyses clearly verify the equiatomic composition of the multicationic oxide nanostructures (Figs. S6 and S7).

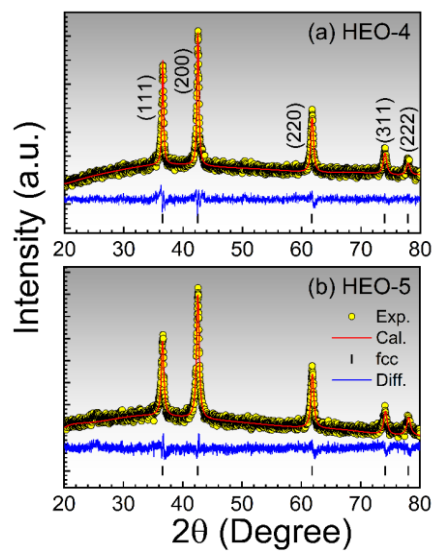

**Fig. 1** Rietveld refinements of the room temperature powder XRD spectra for (a) **HEO-4** and (b) **HEO-5** nanostructures. Experimental data points, theoretical fits according to the *fcc* crystal structure and Bragg positions are represented using yellow circles, solid red lines and black bars, respectively. The residuals are plotted at the bottom in blue color.

Infrared (IR) spectroscopy was used to analyse functional groups present in **HEO-4** and **HEO-5** (Fig. S8). The IR band near 500 cm$^{-1}$ can be assigned to metal-oxygen stretching vibrations, indicating the formation of the oxide phase. The bands observed at 1024 cm$^{-1}$ and 1270 cm$^{-1}$ correspond to the C-N stretching and COO$^-$ bending vibrations of glycine, respectively. The peak at 1630 cm$^{-1}$ has been ascribed to carboxyl stretching vibrations, while broad signals around 3268-3440 cm$^{-1}$ arise from O-H and N-H stretching vibrational mode. This indicates the successful incorporation of glycine-derived functionalities during synthesis,[33] further supporting the formation of multicomponent high-entropy oxide nanoparticles.

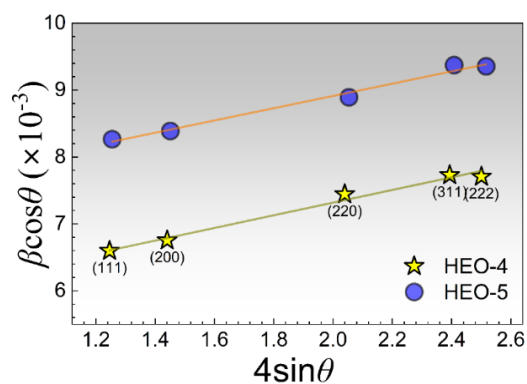

**Fig. 2** Willson-Hall (W-H) plot showing variation of βcosθ as a function of 4sinθ for **HEO-4** (yellow stars) and **HEO-5** (blue circles) nanocatalysts.

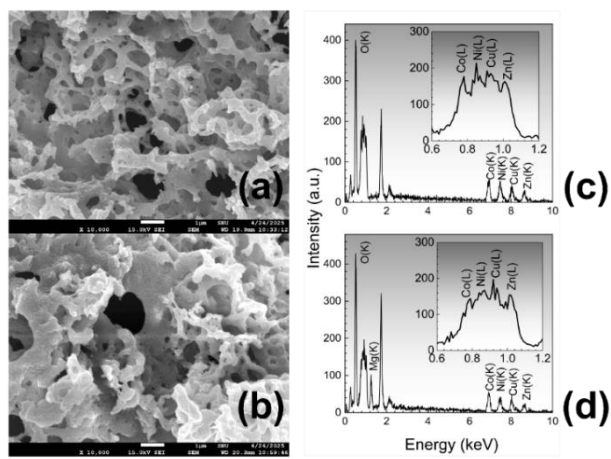

**Fig. 3** FE-SEM images and EDAX spectra of (a, c) **HEO-4** and (b, d) **HEO-5**.

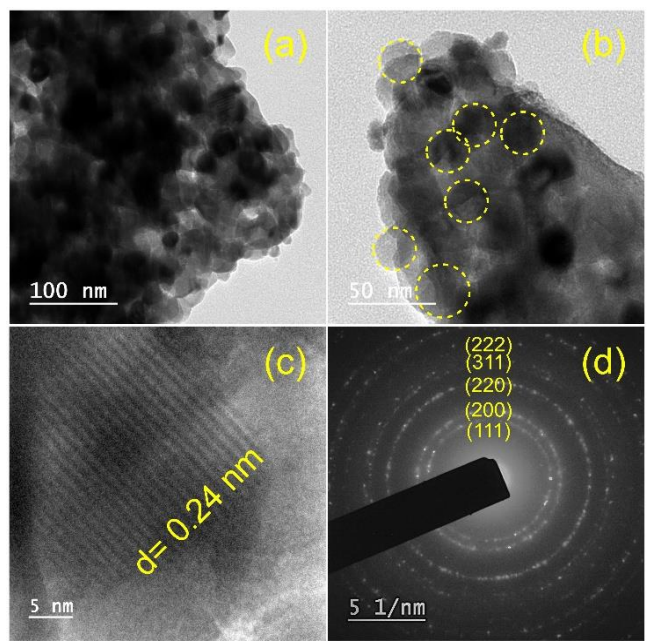

**Fig. 4** (a-b) TEM micrograph for **HEO-5** nanostructures showing agglomerated nearly circular particles; (c) high resolution TEM displaying lattice fringes and (d) SAED pattern confirming *fcc* crystal of these nanostructures.

The porous structures of the synthesized HEOs were examined by nitrogen adsorption-desorption isotherm measurements. The isotherm plots of **HEO-5** and **HEO-4** are presented in Fig. S9. Both high entropy materials exhibit a type-II hysteresis loop, indicative of mesoporous characteristics. The specific surface areas of **HEO-5** and **HEO-4** were estimated to be ≈2.53 and ≈2.25 m$^2$/g, respectively. Pore size distributions, derived from the desorption branch using the Barrett-Joyner-Halenda (BJH) method, revealed average pore diameters of ≈25.3 nm for **HEO-5** and ≈47.4 nm for **HEO-4** which could be attributed to the inter-particle voids among the aggregated nanocrystallites. Such variations in surface area and porous structure can be attributed to entropy stabilization effects and the inherent structural disorder of high-entropy oxides.

The electronic states and surface composition of the HEOs were investigated using XPS analysis. The survey spectra, shown in Fig. S10, confirmed the presence of all the constituent metal ions along with carbon and nitrogen atoms in both the multicationic oxides (**HEO-5** and **HEO-4**). High-resolution XPS spectra were deconvoluted using XPSPEAK41 software. To analyse the surface oxygen vacancy features, the high-resolution O 1s spectra of both the HEOs were deconvoluted and two distinct components were identified at binding energies at 530.7 eV and 532.8 eV, corresponding to lattice-oxygen ($O_L$) and surface-oxygen ($O_S$) respectively.[43] The surface oxygen vacancy was estimated from the [$O_S/O_L$] ratio, which was higher in **HEO-5** (4.83) when compared with **HEO-4** (2.37), indicating greater density of oxygen vacancies in **HEO-5**. This enhanced vacancy formation is likely associated with the incorporation of Mg$^{2+}$, which has a relatively smaller ionic radius, leading to lattice distortion and defect generation.[44] Furthermore, the deconvolution of the Zn 2p spectrum exhibited two spin-orbit component corresponding to Zn $2p_{3/2}$ and Zn $2p_{1/2}$, characteristics of $Zn^{2+}$ species. The Ni 2p spectrum was deconvoluted into its spin-orbit doublets along with satellite peaks, confirming the presence of Ni in 2+ oxidation states. In case of Cu2p, the spectrum was resolved into four spin-orbit peaks along with satellite peaks, indicating the presence Cu in mixed 1+/2+ states. Similarly, the deconvolution of Co 2p spectrum revealed four peaks with satellite features, where the deconvoluted peaks 776eV and 780 eV could be attributed to the Co(II) and Co(III) species, respectively.[45]

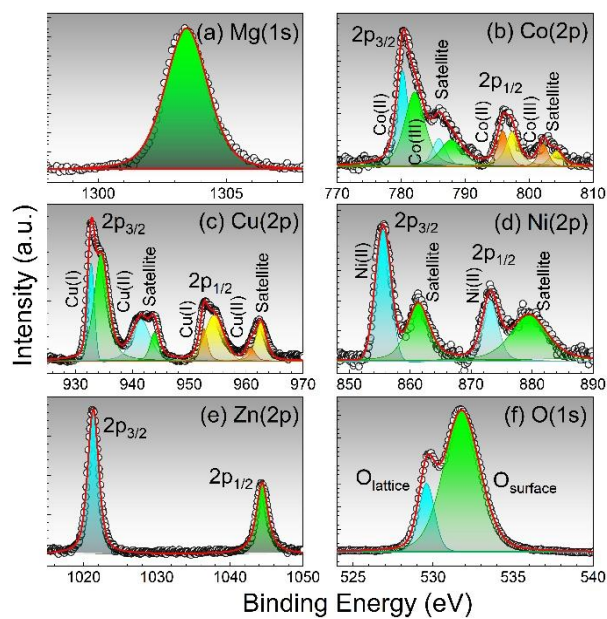

**Fig. 5** Deconvoluted XPS spectra of (a) Mg 1s, (b) Co 2p, (c) Cu 2p, (d) Ni 2p, (e) Zn 2p and (f) O 1s for **HEO-5** nanostructures after background subtraction.

**Catalytic activity**

The as-prepared HEOs were evaluated as catalysts for the hydrogenation of various nitrophenol derivatives using $NaBH_4$ as the reducing agent in aqueous solution. Fig. S11 shows the UV-Vis spectra of *p*-nitrophenol (*p*-NP) in the absence and presence of $NaBH_4$. Upon addition of $NaBH_4$, the solution turned yellow, and the absorption maximum shifted from 320 nm to 400 nm, indicating the formation of the nitrophenolate ion. The progress of the catalytic hydrogenation was monitored by time-dependent UV-Vis spectroscopy at regular intervals. A gradual decrease in the 400 nm band accompanied by the emergence of a new band at 300 nm, along with a distinct isosbestic point. This indicated a stoichiometric transformation of the nitro group to the corresponding amine without any byproducts. These bands correspond to π–π* transitions of nitrophenolate (400 nm) and aminophenolate (300 nm), respectively. Notably, **HEO-5** exhibited significantly faster catalytic activity, achieving full conversion in 2 min compared to ~5 min with **HEO-4**. A similar catalytic trend was observed for other nitrophenol derivatives, including *o*-nitrophenol (*o*-NP), *m*-nitrophenol (*m*-NP), 2,4-dinitrophenol (DNP) and picric acid (PA), further demonstrating the superior catalytic activity of **HEO-5** compared to **HEO-4** as shown in Figs. 7 and S12.

**Kinetic studies**

The kinetics of reduction reaction of different nitrophenol derivatives were analysed by applying pseudo first order rate equation, as $NaBH_4$ was used in large excess.[46,47] The apparent rate constants were calculated using the equation $\ln(A_0/A) = k_{app}\,t$ where $A_0$ and A represent the absorbance of nitrophenolate at the initial time and at time t, respectively, measured at 400 nm. Since the nitrophenol concentration was relatively low, absorbance values could be directly correlated with concentration. Accordingly, linear plots of $\ln(C_0/C)$ *vs* t (Figs. 8a and S13) were obtained for all nitrophenol derivatives, from which rate constants were calculated from the slope. The $k_{app}$ values for different nitrophenol derivatives under identical reaction conditions, using **HEO-5** and **HEO-4** as catalysts, are summarized in Table 1. Notably, higher rate constants were observed for *p*-NP and picric acid compared to *m*-NP and *o*-NP when **HEO-5** was used. This behavior can be attributed to the stronger mesomeric electron-donating effect of the para -OH substituent, which facilitates electron transfer from $NaBH_4$ to the -$NO_2$ group.[46] In contrast, *o*-nitrophenols are stabilized by intramolecular hydrogen bonding, resulting in comparatively lower reactivity.[46]

In catalytic reactions, the turnover frequency (TOF) serves as a key parameter to evaluate the comparative activity of the catalyst and was therefore calculated for reduction of different nitrophenol derivatives using both **HEO-4** and **HEO-5** as catalyst. The TOF (mol $g^{-1}\,s^{-1}$) were calculated by using the following formula:

$$\text{TOF} = \frac{n x M}{100 w t}$$

where **n** and M are the initial number of moles and molecular weight of nitrophenol, x is the percentage conversion of nitrophenols, w is the weight of catalyst used, and t is the reaction time (s).[46,48] The calculated TOF values for the reduction of different nitrophenol derivatives with both catalysts are summarized in Table S1. The calculated TOF values were found to be higher for **HEO-5** compared to **HEO-4** for the reduction of different nitrophenols. The catalytic reduction efficiency of **HEO-5** has been compared with the previously reported catalysts and summarized in Table S1.

**Thermodynamic studies**

To gain deeper insight into the kinetics of nitrophenol reduction, key thermodynamic parameters including activation energy ($E_a$), enthalpy of activation ($\Delta H^{\ddagger}$), and entropy of activation ($\Delta S^{\ddagger}$) were calculated. The activation energy was determined by conducting the reduction of *p*-NP at different temperatures using **HEO-4** and **HEO-5** as catalysts (Figs. 8b and S14). The calculated $E_a$ values for *p*-NP reduction over **HEO-5** and **HEO-4** were found to be 29.11 and 30.08 kJ/mol, respectively (Figs. 8c). The relatively low $E_a$ suggest that the high entropy oxide catalyst reduce the energy barrier, thereby enhancing the hydrogenation efficiency of nitrophenols

The enthalpy and entropy of activation were further evaluated using Eyring equation:

$$\ln\frac{k}{T} = \frac{\Delta S^{\ddagger}}{R} + \ln\frac{k_B}{h} - \frac{\Delta H^{\ddagger}}{RT}$$

Gibbs free energy of activation was calculated as:

$$\Delta G^{\ddagger} = \Delta H^{\ddagger} - T\Delta S^{\ddagger}$$

here activation parameters ΔH‡, ΔG‡, ΔS‡ are enthalpy, Gibb's free energy and entropy, respectively, while $k_B$, R and h represent Boltzmann's constant, universal gas constants and Planck's constants, respectively. The activation parameters obtained from the Eyring plots (Fig. 8d) are summarized in Table S2. For **HEO-5**, the negative value of ΔS‡ indicates the formation of an associative transition state, while the positive enthalpy of activation (ΔH‡) indicates the energy requirement for formation of the intermediate.[48] The calculated Gibbs free energy of activation follows the trend ΔG‡(**HEO-5**) < ΔG‡(**HEO-4**), which explain the higher catalytic efficiency observed for **HEO-5**.[48] A summary of the calculated thermodynamic parameters is depicted in Table S2. These results highlight the role of entropy stabilization in driving the catalytic hydrogenation process.

**Table. 1** Apparent rate constants ($k_{app}$) and TOF values for the reduction of different nitrophenol derivatives using **HEO-4** and **HEO-5**.

| Substrate | $k_{app}$ (s$^{-1}$) | | TOF (molg$^{-1}$s$^{-1}$) | |
|---|---|---|---|---|
| | HEO-4 | HEO-5 | HEO-4 | HEO-5 |
| *p*-NP | 0.54×10$^{-2}$ | 3.22×10$^{-2}$ | 7.2×10$^{-4}$ | 2.1×10$^{-3}$ |
| *o*-NP | 0.73×10$^{-2}$ | 1.08×10$^{-2}$ | 2.5×10$^{-3}$ | 5.1×10$^{-3}$ |
| *m*-NP | 1.62×10$^{-2}$ | 1.80×10$^{-2}$ | 9.45×10$^{-3}$ | 7.6×10$^{-3}$ |
| 2,4 DNP | 0.85×10$^{-2}$ | 1.07×10$^{-2}$ | 2.23×10$^{-3}$ | 7.09×10$^{-3}$ |
| PA | 1.53×10$^{-2}$ | 3.42×10$^{-2}$ | 5.67×10$^{-3}$ | 8.5×10$^{-3}$ |

Reaction conditions: 0.5 mg catalyst, 0.003 mmol substrate, molar ratio of nitrophenol to NaBH$_4$ is 1:60, RT.

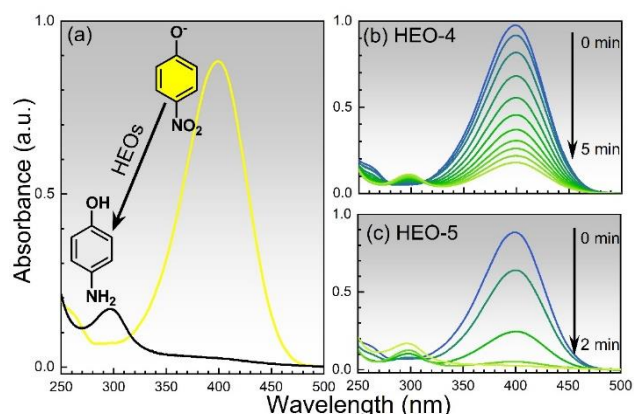

**Fig. 6** (a) UV-Vis spectral changes for the reduction of *p*-nitrophenolate to *p*-aminophenol. Spectral profiles showing catalytic reduction of *p*-NP using 0.5 mg of **HEO-4** (b) and **HEO-5** (c) catalysts, respectively.

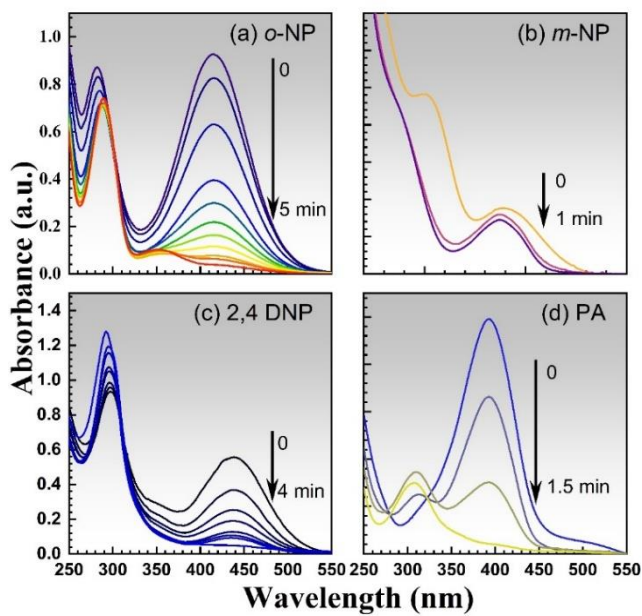

**Fig. 7** Changes in the UV-Vis spectra showing catalytic reduction of (a) *o*-NP, (b) *m*-NP, (c) 2,4 DNP and (d) PA using 0.5 mg of **HEO-5**.

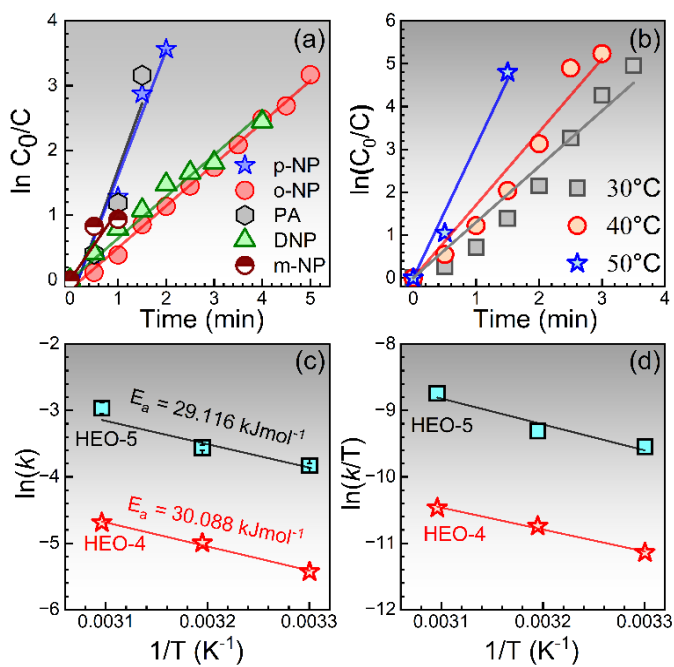

**Fig. 8** (a) ln(C$_0$/C) vs time plot for the catalytic reduction of *p*-NP, *o*-NP and PA using **HEO-5**. (b) Temperature dependent rate constant determination for the catalytic reduction of *p*-NP using **HEO-5** (5 mg) catalysts (c) Arhenius plot to extract activation energy (d) Eyring plot showing linear variation of ln(k/T) as a function of 1/T to extract activation thermodynamic parameters.

**Possible mechanism**

The catalytic reduction of nitrophenol derivatives over **HEO-5** likely proceeds *via* a surface-mediated electron transfer mechanism. In general, such reductions occur *via* either the Langmuir-Hinshelwood (L-H) or the Eley-Rideal (E-R) pathway.[47,49–51] In L-H mechanism, both reactants are simultaneously adsorbed onto the catalyst surface, whereas in the latter, only one reactant is adsorbed and reacts directly with the other in solution.[50] In order to confirm the mechanistic pathway, kinetic studies were conducted by varying the concentration of *p*-NP while keeping NaBH$_4$ constant, and vice versa. Notably, the first order rate constant ($k_{app}$) decreased with increasing concentration of *p*-NP (Fig. S15). On the other hand, a significant improvement in $k_{app}$ could be realized with increasing NaBH$_4$ concentration (Fig. S16). These non-linear variations of rate constant confirm that the reduction process follows L-H mechanism. Following this mechanistic pathway, NaBH$_4$ first readily adsorbs onto of the **HEO-5** surface and transfers hydride species to the active site of the catalyst. Simultaneously, nitrophenols are adsorbed through interaction of the -NO$_2$ group mostly with surface oxygen vacancies.[52–55] The high configurational entropy and multicomponent nature of **HEO-5** generate abundant oxygen vacancies and heterometallic active centres, as evidenced by XPS analysis. These defect sites facilitate strong adsorption and activation of the –NO$_2$ group of nitrophenols through electron-deficient oxygen atoms, promoting electron transfer. The close proximity of both the adsorbed reactants on the catalytic surface facilitates rapid hydride transfer from BH$_4^-$ to the adsorbed nitro group. This process reduces nitro to amine *via* the formation of nitroso (-NO) and hydroxylamine (-NHOH) intermediates (Scheme 2).[56] Eventually, the aminophenol products desorbed from the surface and new catalytic cycle takes place. Furthermore, the synergistic interaction of metal ions in HEO-5 offers a range of redox active sites which enhances charge delocalization and accelerate electron transfer leading to enhanced catalytic efficacy. The enhanced defect density and heterometallic catalytically active sites lower the activation barrier for hydride transfer and stabilization of the reaction intermediates. This is supported by calculated low activation energy and negative activation entropy of **HEO-5**, enabling a fast and efficient hydrogenation reaction.

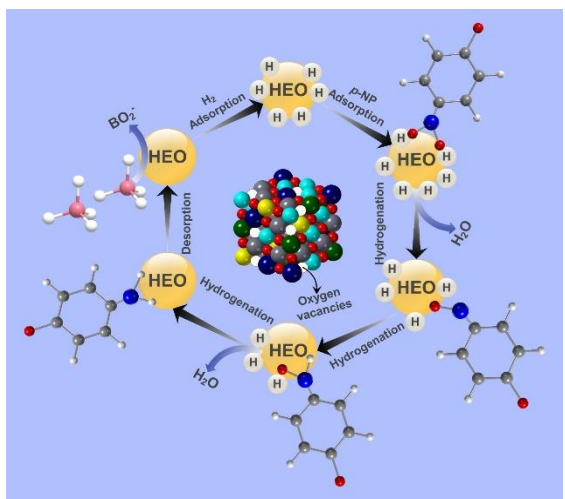

**Scheme 2** A schematic showing the mechanism of catalytic reduction of nitrophenol.

**Stability and scalability tests of HEO-5.** For sustainable industrial amine production, catalysts must retain high activity while withstanding harsh reaction conditions. In this study, the entropy-stabilized metal oxide phase of **HEO-5**, a homogeneous solid solution of five binary oxides, demonstrated excellent structural stability and served as an efficient catalyst for the hydrogenation reaction.

The long-term stability and reusability of **HEO-5** were evaluated through consecutive recycling experiments for the reduction of *p*-nitrophenol. The catalyst retained its activity over five successive cycles, with negligible loss in conversion efficiency as shown in Fig. S17. To further assess structural robustness, reduction reaction was performed at a higher substrate concentration, where **HEO-5** continued to exhibit efficient catalytic performance. The PXRD analysis of the recovered catalyst after the reduction reaction revealed negligible change in diffraction patterns, confirming that the crystalline phase and structural integrity of **HEO-5** remained unaltered after repeated use Fig. S18. These results demonstrate the structural robustness and scalability of **HEO-5**, highlighting its potential for sustainable catalytic applications.

## Conclusions

Two multi-cationic oxide nanostructures namely **HEO-4** and **HEO-5**, having theoretical configuration entropy in the region of medium and high entropy values, were successfully synthesized to compare the effect of entropy on their catalytic activity. Single phase with face-centered cubic crystal structure (space group *Fm3m*) was confirmed through Rietveld refinements on the powder x-ray diffraction data. Nearly spherical shaped particles with an average diameter of ~19 nm were evident from electron microscopy. The highly porous nature of these nanostructures, confirmed by SEM and BET analyses, facilitated superior catalyst-analyte interactions, thereby enhancing the catalytic conversion of various nitrophenol derivatives, including *p*-NP, *o*-NP, *m*-NP, DNP, and PA. The reaction mechanism was further elucidated through XPS, kinetic, and thermodynamic studies. Overall, this study highlights the crucial role of configurational entropy and oxygen vacancies in regulating catalytic efficiency, thereby paving the way for the development of next-generation high-entropy oxide nanocatalysts for environmental remediation and large-scale industrial chemical conversions.

## Conflicts of interest

"There are no conflicts to declare".

## Acknowledgements

A. V. gratefully acknowledges the PhD fellowship support received from UPES, Dehradun. S. L., T.G. and S.K. acknowledges the research funding support provided by UPES through the SEED program (Grant No.: UPES/R&D-SOE/20062022/02) and SEED-INFRA initiatives (Grant Nos.: UPES/R&D-SEED-INFRA/24082023/04 and UPES/R&D-SoAE/08042024/20). Part of the experimental work was carried out utilizing the advanced research facilities at the Central Instrumentation Centre (CIC), UPES, Dehradun.